\documentclass[aps,prl,twocolumn,showpacs,groupedaddress]{revtex4}
\usepackage{graphicx} 
\usepackage{amsmath}
\usepackage{longtable}

\begin{document}

\title{Second bound state of PsH}
\author{J.Mitroy}
  \email{jxm107@rsphysse.anu.edu.au}
\affiliation{Faculty of Technology, Charles Darwin University, Darwin NT 0909, Australia}
\author{M.W.J.Bromley}
  \email{mbromley@physics.sdsu.edu}
\affiliation{Department of Physics, San Diego State University, San Diego CA 92182, USA}

\date{\today}

\begin{abstract}

The existence of a second bound state of PsH that is electronically 
stable and also stable against positron annihilation by the normal 
$2\gamma$ and $3\gamma$ processes is demonstrated by explicit calculation.  
The state can be found in the $^{2,4}$S$^{\rm o}$ symmetries with 
the two electrons in a spin triplet state.  The binding energy
against dissociation into the H($2p$) + Ps($2p$) channel was 
$6.06 \times 10^{-4}$ Hartree.  The dominant decay mode of the states 
will be radiative decay into a configuration that autoionizes or 
undergoes positron annihilation.  The NaPs system of the same 
symmetry is also electronically stable with a binding 
energy of $1.553 \times 10^{-3}$ Hartree with respect to the Na($3p$) 
+ Ps($2p$) channel.

\end{abstract}

\pacs{36.10.-k, 36.10.Dr, 34.85.+x}

\maketitle 

The stability of a bound state composed of two electrons and a
positron, the positronium negative ion, was first demonstrated
in a seminal calculation by Wheeler \cite{wheeler46}.  Shortly 
after this calculation, the four body systems, PsH and Ps$_2$ 
were shown to be stable \cite{hylleraas47b,ore51}.  Since that
time, only a few other electronically stable states have been
discovered that can be formed from combinations of $p^+$,$e^-$ 
and $e^+$.  These are additional bound states of Ps$_2$ 
\cite{kinghorn93,varga98a,usukura00a}, a compound that is best 
described as $e^+$PsH \cite{varga99a}, and a ($p^+$, 4$e^-$, 2$e^+$) 
complex \cite{varga99a}.  Additionally, a number of atoms 
have been identified as being capable of binding positronium
and positrons \cite{ryzhikh97a,strasburger98,mitroy02b}      

A common feature of all these systems is that the positron
annihilation process occurs by either a $2\gamma$ or $3\gamma$ 
process with rates of order 10$^9$ s$^{-1}$ or 10$^{6}$ s$^{-1}$
(for those systems for which an annihilation rate has been 
determined).  In the present letter, we identify a new class of 
positronic compounds that are electronically stable, and in 
addition they have the unusual feature of decaying very slowly by 
$2\gamma$ or $3\gamma$ annihilation. Stable variants of PsH, 
and NaPs are identified and initial estimates of their binding 
energies are given.  The existence of a new bound state of PsH 
is somewhat surprising given the amount of activity involved in 
identifying the resonant states of the Ps-H complex 
\cite{drachman79b,dirienzi02both,yan03a}.
The new PsH and NaPs bound states are unnatural parity 
states with symmetry conditions that act to prevent positron 
annihilation and to also prevent decay into the the lowest 
energy dissociation products.  These systems have the two 
electrons in a spin triplet state, a total orbital angular momentum 
of zero, and an odd parity, i.e. $L^{\Pi} = 0^-$.  Positron 
annihilation by the $2\gamma$ or $3\gamma$ process is
forbidden for such a state.  

First consider the $2\gamma$ process (which occurs at a rate 
of $8 \times 10^9$ s$^{-1}$ for the Ps ground state).  For 
this process to occur, the annihilating electron-positron 
pair must be in a spin singlet state and the relative angular 
momentum must be zero.  (The decay 
rate is not absolutely zero since the Ps($2p$) levels can undergo 
$2\gamma$ and $3\gamma$ annihilation at rate proportional to 
$\alpha^5$ and $\alpha^6$ respectively \cite{alekseev58a,alekseev58b}.  
The rates for the different Ps($2p$) levels have been calculated 
to be approximately 10$^4$ s$^{-1}$ \cite{alekseev58a,alekseev58b}.)   

Now consider the electron-positron annihilation of a PsH state 
of $^{2}$S$^{\rm o}$ symmetry.  The relative angular momentum of
the annihilating pair ($L_{\rm rel}$) must be zero.   This means 
the total angular momentum of the state will come from 
the center-of-mass motion of the annihilating pair ($L_{\rm cm}$), 
and from the angular momentum of the spectator electron 
($L_{\rm spectator}$).   The total parity of the state is 
determined by the parity of the individual constituents, i.e.
$\Pi = (-1)^{L_{\rm spectator}+L_{\rm cm}+L_{\rm rel}}$.  It 
is simply not possible to form an odd parity state with a total 
angular momentum of zero if any one of the constituent angular 
momenta is zero.  Consequently, a two electron/one positron 
state of $^{2}$S$^{\rm o}$ symmetry cannot decay by 
the fast $2\gamma$ process.

These arguments also apply to the $3\gamma$ annihilation
process.   The $3\gamma$ process occurs for electron-positron
pairs in a spin triplet state with a relative angular momentum
of zero.  Once again, it is simply impossible to form a state
of $^{2}$S$^{\rm o}$ (or $^{4}$S$^{\rm o}$) symmetry if the
relative angular momentum of the annihilating pair is zero.
So it is reasonable to conclude that the lowest order $3\gamma$ 
decay is not possible from a $^{2,4}$S$^o$ state.  

These $L^{\Pi}$ conditions also act to prevent the dissociation 
of these four-body systems into combinations of the lower energy
dissociation channels.  Once again consider a $^{2}$S$^{\rm o}$ 
state of PsH. Dissociation into Ps($1s$)+H($1s$) is forbidden 
since $\Pi = (-1)^L$ where $L$ is the orbital angular momentum
between the Ps($1s$) and H($1s$) fragments.  Similarly, dissociation
into Ps($ns$)+H($n\ell$) or Ps($n\ell$)+H($ns$) does not occur 
since it is not possible 
to construct an $L^{\Pi} = 0^{-}$ state if one of the angular 
momentum is zero.  The lowest energy dissociation channel would 
be into Ps($2p$)+H($2p$) ($p$-wave) with an energy of $-0.1875$ 
Hartree.  Another possible decay would be into the 
H$^-$(2p$^2$ $^3$P$^e$)+$e^+$ channel with a threshold energy 
of $-0.125355$ Hartree \cite{holoien61a,drake70a,bylicki03a}.  The 
stability of the H$^-$($2p^2$ $^3$P$^{\rm e}$) bound state also 
suggests a mechanism for binding.  One can think of the positron 
trapped into a $2p$ state of the H$^-$ attractive potential well.  
If the H$^-$ state is regarded as a point particle with an internal  
energy of $\approx -0.125$ Hartree, then a positron in the $2p$ 
state will lower the total energy to $-0.250$ Hartree.  In
actuality the H$^-$($2p^2$ $^3$P$^{\rm e}$) state is very 
diffuse, but this model does suggest that there is a large 
energy advantage associated with binding the positron to
the negative ion.   

All the calculations in the present paper were performed with
a configuration interaction approach 
\cite{bromley02a,bromley02b,mitroy06a}.  The CI basis was 
constructed by letting the two electrons (particles 1 and 2) 
and the positron (particle 0) form all the possible total 
angular momentum $L_T = 0$ configurations, with the two 
electrons in a spin-triplet state, subject to the selection rules,
\begin{eqnarray}
\max(\ell_0,\ell_1,\ell_2) & \le & J \ , \\
\min(\ell_1,\ell_2)& \le & L_{\rm int} \ ,  \\  
(-1)^{(\ell_0+\ell_1+\ell_2)}& = & -1  \ . 
\end{eqnarray}
In these rules $\ell_0$, $\ell_1$ and $\ell_2$ are respectively 
the orbital angular momenta of the positron and the two electrons.  
We define $\langle E \rangle_J$ to be the energy of the calculation 
with a maximum orbital angular momentum of $J$.  The single
particle orbitals were Laguerre Type Orbitals (LTOs) with a 
common exponent chosen for all the orbitals of a common $\ell$  
\cite{bromley02a,bromley02b,mitroy06a}.   
The orbitals basis sets for the positron and electrons were 
identical.      

\begin{table}[th]
\caption[]{  \label{tab1}
The energy of the $^{2,4}$S$^{\rm o}$ state of PsH as a function 
of $J$.  The threshold for binding is $-0.1875$ Hartree. The column
$n$ gives the total number of occupied electron orbitals 
(the number of positron orbitals was the name) while $N_{CI}$
gives the total number of configurations.  The results of the
$J \to \infty$ energy extrapolations at $J = 10$ are also given.
}
\begin{ruledtabular}
\begin{tabular}{lccc} 
$J$ &  $n$   &   $N_{CI}$ & $\langle E \rangle_{J}$   \\ \hline
 1  &   15   &    1800  & $-0.16755817$  \\ 
 2  &   30   &    6975  & $-0.17938456$  \\ 
 3  &   45   &   19125  & $-0.18327387$  \\ 
 4  &   60   &   36000  & $-0.18510508$  \\ 
 5  &   75   &   54675  & $-0.18612672$  \\ 
 6  &   90   &   74925  & $-0.18675211$  \\
 7  &  105   &   95175  & $-0.18715811$  \\
 8  &  120   &  115425  & $-0.18743280$  \\
 9  &  135   &  135675  & $-0.18762315$  \\
10  &  150   &  155925  & $-0.18775631$  \\
\hline  
 &   &   &  \multicolumn{1}{c}{$\langle E \rangle_{\infty}$}      \\
\multicolumn{3}{l}{1-term eq.(\ref{extrap1})}   & $-0.18797567$   \\
\multicolumn{3}{l}{2-term eq.(\ref{extrap1})}   & $-0.18806917$   \\
\multicolumn{3}{l}{3-term eq.(\ref{extrap1})}   & $-0.18810659$   \\
\end{tabular}
\end{ruledtabular}
\end{table} 

The Hamiltonian was diagonalized in a basis constructed from a 
large number of single particle orbitals, including orbitals up 
to $\ell = 10$.  There were $15$ radial basis functions for each 
$\ell$.  Note, the symmetry of the state prevented the electrons
or positrons from occupying $\ell = 0$ orbitals.  The largest 
calculation was performed with $J = 10$ and
$L_{\rm int} = 3$ and gave a CI basis dimension of 155925.  
The parameter $L_{\rm int}$ does not have to be particularly large
since it is mainly concerned with electron-electron correlations 
\cite{bromley02b}.  The resulting Hamiltonian matrix was diagonalized 
with the Davidson algorithm \cite{stathopolous94a}, and a total 
of 300 iterations were required for the largest calculation.   

The energy of the PsH $^{2,4}$S$^o$ state as a function of $J$ is
given in Table \ref{tab1}. The calculations only give an energy 
lower than the H($2p$) + Ps($2p$) threshold of $-0.1875$ Hartree 
for $J \ge 9$.  A major technical problem afflicting CI calculations 
of positron-atom interactions is the slow convergence of the energy 
with $J$ \cite{mitroy02b,mitroy06a}.  The  $J \rightarrow \infty$ 
energy, $\langle E \rangle_{\infty}$, is determined by the use of 
an asymptotic analysis.  The successive increments, 
$\Delta E_{J} = \langle E \rangle_J - \langle E \rangle_{J-1}$, 
to the energy can written as an inverse power series 
\cite{schwartz62a,carroll79a,hill85a,mitroy06a,bromley06a}, viz 
\begin{equation}
\Delta E_J \approx \frac {A_E}{(J+{\scriptstyle \frac{1}{2}})^6} 
    + \frac {B_E}{(J+{\scriptstyle \frac{1}{2}})^7} 
    + \frac {C_E}{(J+{\scriptstyle \frac{1}{2}})^8} + \dots \ \   .
\label{extrap1}
\end{equation}
The first term in the series starts with a power of 6 since
all the possible couplings of any two of the particles result
in unnatural parity states \cite{kutzelnigg92a}.    

\begin{figure}[tbh]
\centering{
\includegraphics[width=9.0cm,angle=0]{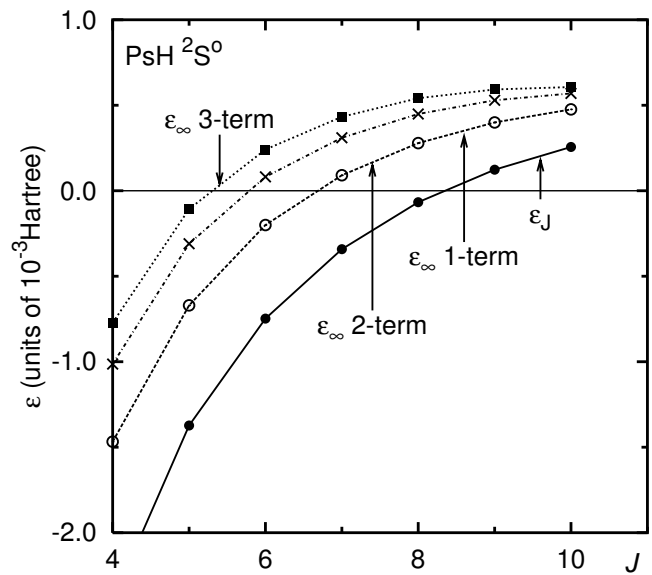}
}
\caption[]{ \label{fig:PsHE}
The binding energy, $\varepsilon = -(\langle E \rangle+0.1875)$,  
of the $^{2,4}$S$^{\rm e}$ state of PsH as a function of $J$.  
The directly calculated energy  
is shown as the solid line while the $J \to \infty$ limits using 
eq.~(\ref{extrap1}) with 1, 2 or 3 terms are shown as the dashed 
lines.  The H($2p$) + Ps($2p$) dissociation threshold is shown 
as the horizontal solid line.
}
\end{figure}

The $J \to \infty$ limit, has been determined by fitting sets of 
$\langle E \rangle_J$ values to asymptotic series with either 1, 2 
or 3 terms.  The coefficients, $A_E$, $B_E$ and $C_E$ for the 3-term 
expansion are determined at a particular $J$ from 4 successive energies
($\langle E \rangle_{J-3}$, $\langle E\rangle_{J-2}$, 
$\langle E \rangle_{J-1}$ and $\langle E \rangle_{J}$).  
Once the coefficients have been determined it
is easy to sum the series to $\infty$ and obtain the 
variational limit.   Application of asymptotic series analysis
to helium has resulted in CI calculations reproducing the ground
state energy to an accuracy of $\approx \!\! 10^{-8}$ Hartree
\cite{salomonson89b,bromley06a}.  

Figure \ref{fig:PsHE} shows the estimates of 
$\langle E \rangle_{\infty}$ as a function of $J$.  A quick
visual examination suggests that the extrapolations are
converging to a common energy while the energy of 
the three-term extrapolation is close to stabilized at 
$J = 10$.  The impact of the extrapolations is significant 
since they more than double the binding energy.  
The most precise estimate of the binding energy is the 
three-term extrapolation at $J = 10$, namely 
$6.06 \times 10^{-4}$ Hartree (this is computed using 
the $\langle E \rangle_{\infty}$ in Table \ref{tab1}).

\begin{table}[th]
\caption[]{  \label{tab2}
The energy of the $^{2,4}$S$^{\rm o}$ state of NaPs as a function 
of $J$.  The threshold for binding is $-0.17410932$ Hartree. The 
column $n_-$ gives the total number of occupied electron orbitals, 
$n_+$ gives the number of positron orbitals,  while $N_{CI}$
gives the total number of configurations.  The results of the
$J \to \infty$ energy extrapolations at $J = 10$ are also given.
}
\begin{ruledtabular}
\begin{tabular}{lccccc} 
$J$ &  $n_-$ & $n_+$  &   $N_{CI}$ & $\langle E \rangle_{J}$ &  $\varepsilon_J$   \\ \hline
 1 & 16  &  15 &   2040  & $-0.153740547$ & $-0.02031794$  \\  
 2 & 31  &  30 &   7440  & $-0.166301513$ & $-0.00775597$  \\  
 3 & 46  &  45 &  19815  & $-0.170503790$ & $-0.00355470$  \\ 
 4 & 61  &  60 &  36915  & $-0.172480804$ & $-0.00157768$  \\  
 5 & 76  &  75 &  55815  & $-0.173572387$ & $-0.00048610$  \\  
 6 & 91  &  90 &  76290  & $-0.174230717$ & 0.00017223  \\   
 7 & 106 & 105 &  96765  & $-0.174650110$ & 0.00059162 \\   
 8 & 121 & 120 & 117240  & $-0.174928718$ & 0.00087023  \\   
 9 & 136 & 135 & 137715  & $-0.175119526$ & 0.00106104  \\   
10 & 151 & 150 & 158190  & $-0.175253262$ & 0.00119477  \\ \hline  
 &   &   &  & \multicolumn{1}{c}{$\langle E \rangle_{\infty}$}  &  \multicolumn{1}{c}{$\varepsilon_{\infty}$}      \\
\multicolumn{4}{l}{1-term eq.(\ref{extrap1})}   & $-0.17547356$ & 0.00141507 \\
\multicolumn{4}{l}{2-term eq.(\ref{extrap1})}   & $-0.17556808$ & 0.00150960 \\ 
\multicolumn{4}{l}{3-term eq.(\ref{extrap1})}   & $-0.17561102$ & 0.00155252  \\
\end{tabular}
\end{ruledtabular}
\end{table} 

Having the established the stability of the $^{2,4}$S$^{\rm o}$ 
state of PsH, it is natural to ask whether other systems with 
this symmetry are stable.  The obvious candidates are the
alkali atoms, since some of them have $np^2$ $^3$P$^{\rm e}$ 
negative ion bound states \cite{norcross74a} that can 
act as a parent state to bind the positron.  The treatment of such 
systems requires the use of a frozen core approximation.  The
details of this approximation have been discussed in great
detail elsewhere \cite{bromley02a,bromley02b,mitroy06a}, 
so only the briefest description is given here.  The model 
Hamiltonian is initially based on a Hartree-Fock (HF) wave 
function for the neutral atom ground state.  The core orbitals
are then frozen.  The direct part of the core potential is 
attractive for electrons and repulsive for the positron.
The impact of the direct and exchange part of the HF core
interactions on the active particles are computed without 
approximation.  One- and two-body semi-empirical polarization 
potentials are then added to the potential.  The adjustable 
parameters of the core-polarization potential are defined by 
reference to the spectrum of neutral atom 
\cite{bromley02b,mitroy03f}.

The system that was investigated was the $^{2,4}$S$^{\rm o}$ 
state of NaPs.  The energies of the $3s$ and $3p$ states in the 
model potential were $-0.18885491$ and $-0.11156294$ Hartree.  The
experimental binding energies are $-0.188858$ and $-0.111547$   
Hartree respectively \cite{nistasd3}.  Electronic stability 
requires a total 3-body energy of $-0.17405849$ Hartree.  
The energy of the $^3$P$^{\rm e}$ excited state of Na$^-$
is $-0.11342529$ Hartree, i.e the Na($3p$) has an electron affinity
of 0.001862 Hartree with respect to attaching an electron to
the $^3$P$^{\rm e}$ state.  This is reasonably close to the
original value of Norcross, 0.00228 Hartree \cite{norcross74a}.

\begin{figure}[bht]
\centering{
\includegraphics[width=9.0cm,angle=0]{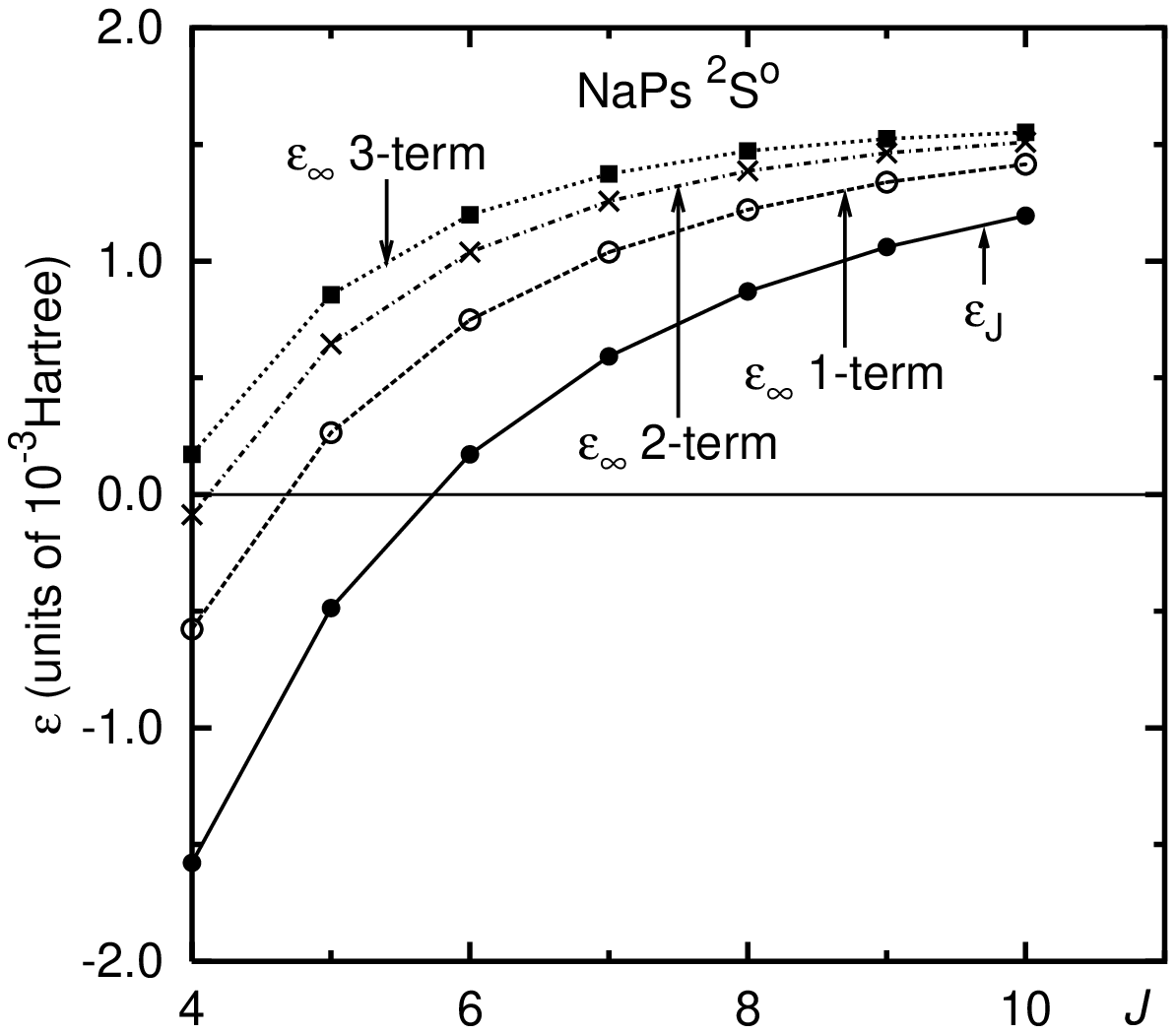}
}
\caption[]{ \label{fig:NaPsE}
The binding energy (in units of Hartree) of the $^{2,4}$S$^{\rm o}$ 
state of NaPs as a function of $J$.  The directly calculated binding 
energy is shown as the solid line while the $J \to \infty$ limits using 
eq.~(\ref{extrap1}) with 1, 2 or 3 terms are shown as the dashed 
lines.  The Na($3p$) + Ps($2p$) dissociation threshold is shown 
as the horizontal solid line.
lines.   
}
\end{figure}

The calculations upon NaPs were very similar in scope and scale
to those carried out upon PsH.  About the only difference was
that an extra $\ell=1$ orbital was added to the electron basis.
Table \ref{tab2} gives the 3-body energy (relative to the Na$^+$
core) as a function of $J$.  The binding energy $\varepsilon_J$
is defined as $\varepsilon_J = -(\langle E \rangle + 0.17410932)$.   
The positron complex is more tightly bound than for PsH and becomes 
electronically stable when $J \ge 5$.

Figure \ref{fig:NaPsE} shows the variation of $\varepsilon_{\infty}$ 
as a function of $J$.  Once again the two and three term extrapolations
seem to be converging to a common energy which is somewhat larger
than the best explicit calculation.  The 3-term value of 
$\varepsilon_{\infty}$ determined at $J = 10$ was 0.001553 Hartree.
This is probably the the best estimate of the binding energy of the 
complex.  The positron can annihilate with the core electrons 
via the $2\gamma$ process since the symmetry considerations are
irrelevant here.  However, the annihilate rate of 
$\Gamma_{\rm core} = 1.66 \times 10^6$ s$^{-1}$ is small because
the positron cannot occupy a $\ell = 0$ orbital.      
  
The PsH and NaPs $^{2,4}$S$^{\text o}$ complexes are stable against
auto-ionizaton, and only decay slowly by positron annihilation.  
However there are other possible decay modes.
Both these complexes can emit a photon, decaying to a state of  
$^{2,4}$P$^{\text e}$ symmetry.  For example, a Ps($np$) fragment in
the complex can emit a photon decaying to a Ps($1s$) type fragment.
The Ps($1s$) fragment could then annihilate by the $2\gamma$ or 
$3\gamma$ process.  In addition, a $^{2,4}$P$^{\text e}$ state could
also decay by auto-ionization.  Due to their low binding energies, 
these systems can be expected to have a structure composed of an
Ps($2p$) cluster loosely bound to an atomic X($np$) excited state.
The lifetime of these states can be expected to be comparable 
to the lifetime of the fragments against single photon decay, e.g. 
H($2p$) $\to$ H($1s$).  So the overall lifetimes of the states can be
expected to be of order $10^{-8}$ - $10^{-9}$ seconds.    
 
It is possible that there are other positronic complexes of 
$^{2,4}$S$^{\text o}$ symmetry that are bound.  The K$^-$, Rb$^-$ and 
Cs$^-$ ions have all been predicted to have $np^2$ $^3$P$^{\rm e}$    
bound states.  So the existence of a stable $^{2,4}$S$^{\text o}$ 
positronic complex would seem to be highly likely.  It would also 
be interesting to examine Li as this does not appear to have
a bound $^3$P$^{\rm e}$ negative ion \cite{norcross74a}.      

Besides the alkali atoms, another physical system possibly admitting
an unnatural parity bound state would be the di-positronium molecule. 
There have been two attempts to find such a bound state, they
were unsuccessful or inconclusive \cite{suzuki00a,bao03a}.  However, 
the investigation of Bao and Shi showed that a $^1$S$^{\text o}$ state 
was very close to being bound, even if it was not bound \cite{bao03a}.   
This raises the tantalizing possibility that a more exhaustive
calculation might reveal the existence of a Ps$_2$ state that
decayed very slowly by positron annihilation.  Besides the Ps$_2$
molecule itself, there is the possible existence of a new biexciton
excited state \cite{usukura99a}.  The parent Ps$^-$ $^3$P$^{\text e}$ 
ion is known to be stable for certain $m_{e^+}/m_{e^-}$ mass ratios 
\cite{mills81b,bhatia83a}.  In circumstances where the mass ratios make 
the $^3$P$^{\text e}$ state of the charged exciton ($e^-$,$e^-$,$h$) 
state stable, it could be expected that a biexciton state of 
$^{1,3,5}$S$^{\text o}$ symmetry would be electronically 
stable.    

These calculations were performed on Linux clusters hosted at the 
South Australian Partnership for Advanced Computing (SAPAC) and 
SDSU Computational Sciences Research Center, with technical
support given by Grant Ward, Patrick Fitzhenry and Dr James Otto.  
The authors would like to thank Dr D M Schrader and Dr C W Clark 
for interesting and helpful correspondence. 


\end{document}